\documentclass[
    aps,
    prd,
    reprint,
    superscriptaddress,
    amsmath,
    amssymb,
    longbibliography,
    floatfix
]{revtex4-2}

\usepackage[T1]{fontenc}
\usepackage[utf8]{inputenc}
\usepackage{graphicx}
\usepackage{booktabs}
\usepackage{xcolor}

\usepackage{hyperref}

\hypersetup{
    pdftitle={Multimessenger Constraints on Nuclear Acceleration  in Neutrino-Emitting Seyfert Coronae},
    pdfauthor={Chengchao Yuan},
    colorlinks=true,
    linkcolor=blue,
    citecolor=blue,
    urlcolor=blue
}

\graphicspath{{figures/}}
\begin{document}

\title{Multimessenger Constraints on Nuclear Acceleration\\
in Neutrino-Emitting Seyfert Coronae}

\author{Chengchao Yuan}
\email{chengchao.yuan@ulb.be}
\affiliation{
    Service de Physique Th\'eorique, Universit\'e Libre de Bruxelles,
    CP225 Boulevard du Triomphe, 1050 Brussels, Belgium
}

 \date{\today}

\begin{abstract}
High-energy neutrinos from Seyfert galaxies probe particle acceleration near supermassive black holes, but the accelerated nuclear composition remains uncertain.
We connect stochastic acceleration to nuclear and radiative cascades to assess nuclear contributions.
Acceleration and losses limit the energy per nucleon available to nuclear cascades; without secondary reacceleration, fragmentation cannot raise this scale.
For NGC~1068, joint neutrino and gamma-ray constraints depend on the nuclear mixture: helium-dominated injection permits larger nuclear power fractions than the heavier mixtures considered.
For the harder candidate {neutrino} spectrum of NGC~7469, the adopted model yields more energetic nuclear cascades, while the source-specific flux selections allow larger nuclear fractions for the heavier mixtures than in NGC~1068.
These results show that composition constraints depend on nuclear yields as well as acceleration conditions and available power.
\end{abstract}

\maketitle

\raggedbottom

\section{Introduction}
\label{sec:introduction}
The evidence for high-energy (HE) neutrino emission from the Seyfert galaxy NGC~1068 has opened a direct window onto particle acceleration near supermassive black holes~\cite{IceCube2022}.
Its inferred TeV neutrino flux exceeds the corresponding gamma-ray limits~\cite{Acciari2019MAGIC,HESS2026SeyfertStarburst}, favoring a compact production region opaque to the accompanying HE photons~\cite{Murase2022HiddenHearts,Inoue2019Corona}.
Recent IceCube observations of X-ray-bright active galactic nuclei (AGNs) identify additional candidates, including NGC~7469, whose best-fit neutrino spectrum is substantially harder than that of NGC~1068~\cite{IceCube2026AGN}.

Neutrino production in AGN cores was proposed decades ago in models involving accretion shocks and abortive jets~\cite{Stecker1991AGN,AlvarezMuniz2004RadioQuiet}.
In disk-corona models, accelerated protons in the hot plasma surrounding the accretion disk interact with gas and photons to produce pions and neutrinos~\cite{Inoue2020NGC1068,Murase2020HiddenCores,Kheirandish2021Coronae,Murase2026ConvexSpectrum,DasMuraseZhang2026,Eichmann2022NGC1068}.
The accompanying photons and electron--positron pairs initiate electromagnetic (EM) cascades, redistributing the radiation to lower energies~\cite{Fang2022TeVDiffuse}.
A purely leptonic origin via photon--photon muon-pair production has also been proposed~\cite{HooperPlant2023}, but is strongly disfavored for NGC~1068 by EM and energetic constraints~\cite{Das2024Production}.

{Recent studies} invoke stochastic acceleration in strong turbulence~\cite{Fiorillo2024TurbulentCorona,LemoineRieger2025,Yang2025NGC7469,Yuan2026CoupledAcceleration,Wang2026TurbulentCorona,Mbarek2026NonthermalPowerLaws} or magnetic reconnection (e.g., Refs.~\cite{Fiorillo2024Reconnection,Mbarek2024,Comisso2024IonElectronReconnection}) to produce nonthermal protons~\cite{Groselj2026ElectronIonCorona}.
Reconnection-powered models also link neutrino production to secondary pair creation in the corona~\cite{Karavola2025}.
Time-dependent transport calculations capture the feedback of cascade radiation on proton cooling and acceleration~\cite{Yuan2026CoupledAcceleration}, while models that evolve the turbulence account for its damping by accelerated particles~\cite{LeBihan2026}.
Joint neutrino and gamma-ray analyses further constrain coronal properties~\cite{Blanco2025NGC1068,Testagrossa2026CoronalTurbulence,Carpio2026Multimessenger}.
Population studies also suggest that Seyfert galaxies may contribute to the diffuse astrophysical neutrino background~\cite{Padovani2024NGC1068,Padovani2024NonJettedBackground,Neronov2024SeyfertPopulation,Fiorillo2025DiffuseCorona,Murase2026ConvexSpectrum,Saurenhaus2026SeyfertDiffuse,Yang2026DiffuseCorona,Mbarek2026DiffuseCorona}.

{The composition of accelerated particles in AGN coronae is still uncertain.}
Nuclei undergo fragmentation and produce secondary nucleons, modifying both the neutrino yield and the EM cascade.
{Detailed nuclear-cascade studies of gamma-ray bursts~\cite{Boncioli2017NuclearPhysics,Biehl2018NuclearCascade} and blazars~\cite{Rodrigues2018BlazarNuclei} have demonstrated how nuclear interactions shape the escaping composition and neutrino emission.}
In a jet model for NGC~1068, helium photodisintegration followed by neutron beta decay produces neutrinos~\cite{Yasuda2025BetaDecay}.
Recent coronal calculations for NGC~1068 demonstrate composition-dependent multimessenger emission using prescribed injection spectra at fixed energy per nucleon~\cite{Esmaeili2026}.
These calculations leave open which nuclear spectra can be supplied when acceleration competes with losses, escape, and nuclear destruction.
This is important for neutrino production because the characteristic neutrino energy depends on the parent energy per nucleon.
Photodisintegration feeds lighter nuclei and secondary nucleons that approximately retain this energy scale, so fragmentation cannot overcome an energy limitation imposed during acceleration without secondary reacceleration.

{In this work, we use steady spectra from stochastic acceleration, losses, and escape to construct power-normalized injection into nuclear and radiative cascades.
We vary the coronal radius and the fraction of cosmic-ray (CR) power assigned to nuclei from helium to iron for three composition templates.
For NGC~1068, joint neutrino and gamma-ray constraints permit larger nuclear fractions for the helium-dominated Solar template than for heavier mixtures under the adopted acceleration and power prescriptions.
NGC~7469 provides a complementary test at higher neutrino energies.}

\section{Coronal model and transport}
\label{sec:coronal-model}
We adopt a commonly used framework for compact, turbulent Seyfert coronae (e.g., Refs.~\cite{Fiorillo2024TurbulentCorona,Yuan2026CoupledAcceleration}), following the environmental and stochastic-acceleration prescriptions of Ref.~\cite{Yuan2026CoupledAcceleration}.
The corona is a homogeneous sphere of radius $R$ and volume $V=4\pi R^3/3$ around a black hole of mass $M_{\rm BH}=10^7M_7M_\odot$, with gravitational radius $R_g=GM_{\rm BH}/c^2$.
We adopt a Thomson depth of approximately $0.5$~\cite{Ricci2018CoronalProperties,Fiorillo2024TurbulentCorona,Yuan2026CoupledAcceleration}; assuming comparable electron and proton densities gives the thermal proton density $n_{\rm t}=2.5\times10^{10}M_7^{-1}(R/20R_g)^{-1}\,\mathrm{cm^{-3}}$.
The magnetic-field strength is $B=(4\pi\sigma_B n_{\rm t}m_pc^2)^{1/2}$, where $m_p$ is the proton mass; we fix the magnetic and turbulent magnetizations to $\sigma_B=\sigma_{\rm turb}=0.1$.
For NGC~1068, we use $M_7=1$, luminosity distance $d_L=10.1$~Mpc~\cite{Tully2008LocalVoid}, and redshift $z=0.0038$.
Energies are in the source frame unless stated otherwise.

Disk optical--ultraviolet (OUV) radiation and coronal X-rays provide fixed targets for nuclear interactions and gamma-ray absorption.
We adopt the NGC~1068 spectra of Ref.~\cite{Fiorillo2024TurbulentCorona}, writing $\mathcal L_X(\epsilon)$ and $\mathcal L_{\rm OUV}(\epsilon)$ for their luminosities per logarithmic interval of photon energy $\epsilon$.
The OUV template adopted there follows Refs.~\cite{Fiorillo2024TurbulentCorona,Marconi2004SED}.
The target-photon number density per unit energy is $n_\gamma(\epsilon)=(R/c)\epsilon^{-2}[\mathcal L_X(\epsilon)/V+\mathcal L_{\rm OUV}(\epsilon)/V_{\rm OUV}]$.
The X-ray luminosity is normalized to $L_X=10^{44}\,\mathrm{erg\,s^{-1}}$ integrated over the tabulated spectrum.
The disk spectrum retains the normalization of the adopted template, whose integrated luminosity is $L_{\rm OUV}$.
Its effective volume is $V_{\rm OUV}=4\pi R_{\rm OUV}^3/3$, with $R_{\rm OUV}=3.7\times10^{14}$~cm estimated from the multitemperature disk profile using a characteristic temperature of $3$~eV and a reference disk luminosity of $2.5\times10^{45}\,\mathrm{erg\,s^{-1}}$~\cite{Yuan2026CoupledAcceleration}.

Magnetic turbulence sets the acceleration timescale and supplies the nonthermal power.
For particle momentum $p$, the momentum-diffusion coefficient is $D_{pp}=p^2/t_{\rm acc}$, with acceleration time $t_{\rm acc}=10l_c/(\sigma_{\rm turb}c)$ and coherence length $l_c=\eta R$, with dimensionless $\eta$~\cite{ComissoSironi2019,Fiorillo2024TurbulentCorona,Yuan2026CoupledAcceleration}.
Charged particles escape by rigidity-dependent diffusion, with a minimum residence time $R/c$.
A fraction $\epsilon_{\rm CR}$ of the magnetic dissipation power $L_{\rm mag}$ feeds nonthermal particles, giving $L_{\rm CR}=\epsilon_{\rm CR}L_{\rm mag}$, where
\begin{align}
&L_{\rm mag}=2.1\times10^{44}
\frac{M_7\sigma_{\rm turb}^{3/2}}
{\eta\max(1,\sqrt{\sigma_B})}\left(\frac{R}{20R_g}\right)
\mathrm{erg\,s^{-1}}.
\label{eq:magnetic-power}
\end{align}
For NGC~1068, we fix $(\eta,\epsilon_{\rm CR})=(0.3,0.3)$ while varying radius and composition, so the available CR power increases as $L_{\rm CR}\propto R$.
Previous coronal models and multimessenger constraints guide the benchmark $(\eta,\epsilon_{\rm CR})$ values for both sources~\cite{Yuan2026CoupledAcceleration,Testagrossa2026CoronalTurbulence}.

{Spectroscopy suggests near-solar Fe abundances in NGC~1068 reflectors~\cite{Bauer2015NGC1068Reflection} and supersolar C and N abundances in NGC~7469 outflows~\cite{Arav2020NGC7469Abundances}.}
{The coronal composition remains uncertain, motivating our three representative templates.}
The primary population comprises protons and the seven nuclear species listed in Table~\ref{tab:composition}, following the species selection of Ref.~\cite{Plotko2025TDE}.
We assign a fraction $f_{\rm nuc}$ of the CR acceleration power to nuclei with mass number $A>1$, including helium:
\begin{align}
L_p&=(1-f_{\rm nuc})L_{\rm CR},
\nonumber\\
L_i&=f_{\rm nuc}w_iL_{\rm CR},
\qquad \sum_{i\ne p}w_i=1.
\label{eq:composition-power}
\end{align}
Here $L_p$ and $L_i$ are the powers assigned to protons and species $i$, respectively, and $w_i$ are the normalized non-proton power weights.
{The Solar and carbon--nitrogen--oxygen (CNO) compositions are based on the solar abundances of Ref.~\cite{Asplund2021SolarAbundances}.
Together with the Fe template, they span a helium-dominated mixture, intermediate-mass nuclei, and iron injection.}
Their weights specify a partition of nonthermal power rather than a measurement of the thermal-gas abundances.

\begin{table}[t]
\caption{Normalized weights $w_i$ within the non-proton acceleration power.
Helium is included in $f_{\rm nuc}$.
Values are rounded.}
\label{tab:composition}
\begin{ruledtabular}
\begin{tabular}{lrrr}
Species & Solar & CNO & Fe\\
\hline
${}^{4}\mathrm{He}$ & 0.956833 & 0 & 0\\
${}^{12}\mathrm{C}$ & 0.010092 & 0.283 & 0\\
${}^{14}\mathrm{N}$ & 0.002760 & 0.077 & 0\\
${}^{16}\mathrm{O}$ & 0.022852 & 0.640 & 0\\
${}^{23}\mathrm{Na}$ & 0.000111 & 0 & 0\\
${}^{28}\mathrm{Si}$ & 0.002642 & 0 & 0\\
${}^{56}\mathrm{Fe}$ & 0.004709 & 0 & 1\\
\end{tabular}
\end{ruledtabular}
\end{table}

For each primary species, we calculate a stationary Fokker--Planck (FP) distribution including acceleration, escape, synchrotron and Bethe--Heitler (BH) losses, interactions with ambient protons, photomeson production, and photodisintegration cooling.
{Daughter-nucleus and secondary-nucleon production is omitted from the FP stage and treated in the subsequent nuclear cascade.}
The resulting differential number density $N_i^{\rm FP}(E)$, defined per unit total particle energy $E$, supplies the shape of the effective injection rate $Q_i^{\rm eff}(E)=C_iN_i^{\rm FP}(E)$,
\begin{align}
&V\int dE\,T_i(E)Q_i^{\rm eff}(E)=L_i,
\label{eq:effective-injection}
\end{align}
where $Q_i^{\rm eff}$ is measured per unit volume, time, and energy, and $T_i=E-m_ic^2$ is the kinetic energy of a particle of mass $m_i$.
We equate the assigned acceleration and effective cascade-injection powers, fixing $C_i$ by power normalization independently of the FP escape rate.
The FP equation, power normalization, and numerical checks are given in Appendix~\ref{app:particle-acceleration}.

The nuclear-transport code \textsc{NuCascade} (Appendix~\ref{app:nuclear-interactions}) evolves this effective injection through photonuclear interactions, decay, continuous losses, and escape, explicitly following surviving primaries, daughter nuclei, and secondary nucleons.
{Gas collisions contribute effective continuous cooling during acceleration and lead to pion production and loss of the parent species.}
{The resulting pion injection rates are passed to \textsc{am}$^3$ (Astrophysical Multi-Messenger Modeling), a time-dependent lepto-hadronic code~\cite{Klinger2024AM3}, to calculate pion and muon cooling and decay, neutrino emission, and EM cascades.}
{We include BH cooling in both transport stages and inject the electron--positron pairs produced by nuclear BH interactions into the radiative cascade.}
In the NGC~1068 cases, heavy-nuclear BH radiation is subdominant, predominantly below $1$~MeV; adding it leaves the neutrino spectrum essentially unchanged.
{We neglect EM-cascade feedback on acceleration and nuclear transport, owing to the dominance of coronal X-rays and disk OUV photons~\cite{Yuan2026CoupledAcceleration}, and turbulence damping by accelerated particles, which can become important for $\epsilon_{\rm CR}\sim1$~\cite{LeBihan2026,Lemoine2024Nonlinear}.}

\section{Physical origin of the composition dependence}
\label{sec:physical-origin}
The spectral cutoffs are jointly determined by stochastic acceleration and interactions, with escape also limiting the residence time.
We illustrate this competition for the NGC~1068 benchmark with $R=30R_g$, $f_{\rm nuc}=0.6$, and the Solar template.
As the proton and iron rates in Fig.~\ref{fig:sm-primary-rates} illustrate, gas collisions provide substantial cooling for iron, and photodisintegration rises rapidly at higher energies; {these processes compete with acceleration to shape the supplied HE tails.}

\begin{figure}[htp]
\includegraphics[width=\columnwidth]{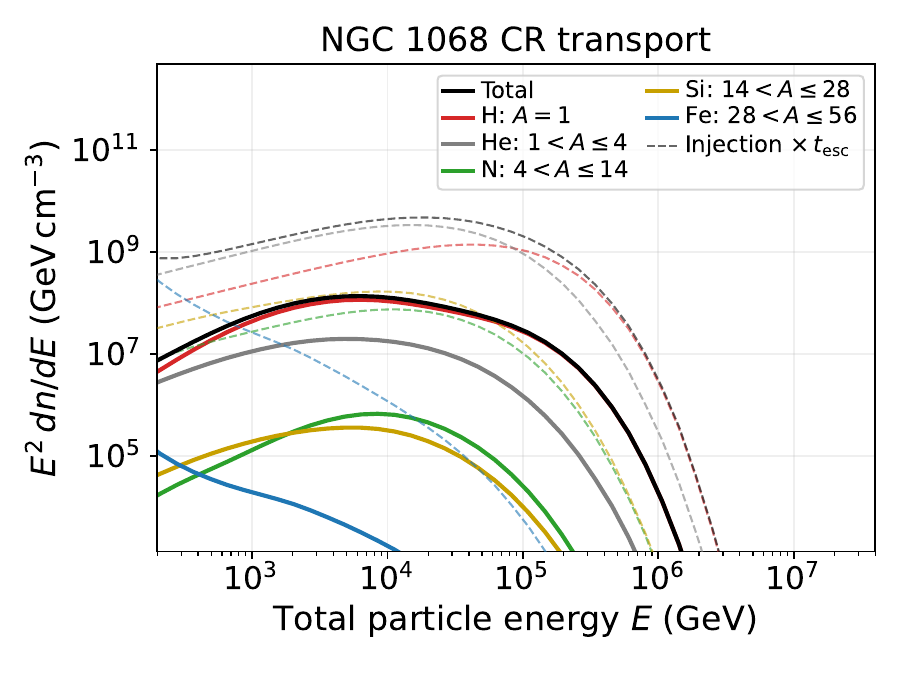}
\caption{{CR densities for the NGC~1068 Solar template at $R=30R_g$ and $f_{\rm nuc}=0.6$.
Solid curves show stationary $E^2N_i(E)$ after transport, including primaries and secondaries; dashed curves show the injection--escape reference $E^2Q_i^{\rm eff}t_{{\rm esc},i}$.
Curves are summed over mass groups (colors); H includes protons and neutrons, and black curves show totals.}}
\label{fig:cr-density}
\end{figure}

Figure~\ref{fig:cr-density} compares the resulting stationary differential number densities $N_i(E)$ (solid curves) with the reference densities $Q_i^{\rm eff}t_{{\rm esc},i}$ (dashed curves), where $t_{{\rm esc},i}$ is the escape time of species $i$.
For the same effective injection, the latter describe the population maintained by injection and escape alone.
{Relative to this reference, continuous cooling shifts particles to lower energies, while interaction and decay sink terms deplete the parent populations.}
Daughter nuclei and secondary nucleons replenish lighter groups, so the change in each group reflects both losses and secondary production.

The characteristic neutrino energy depends on the energy per nucleon, $E_A/A$, for a parent nucleus of total energy $E_A$ and mass number $A$.
At relativistic energies, fragmentation approximately preserves the parent Lorentz factor, giving $E_d/A_d\simeq E_A/A$ for a daughter of total energy $E_d$ and mass number $A_d$.
Without secondary reacceleration, fragmentation redistributes energy among nuclei and free nucleons without increasing their characteristic energy per nucleon.
Figure~\ref{fig:sm-cascade} illustrates the resulting isotope distribution for the same {benchmark with the Solar template}.

\begin{figure*}[t]\centering
\begin{minipage}[t]{0.5\textwidth}
\vspace{0pt}
\centering
\includegraphics[width=\linewidth]{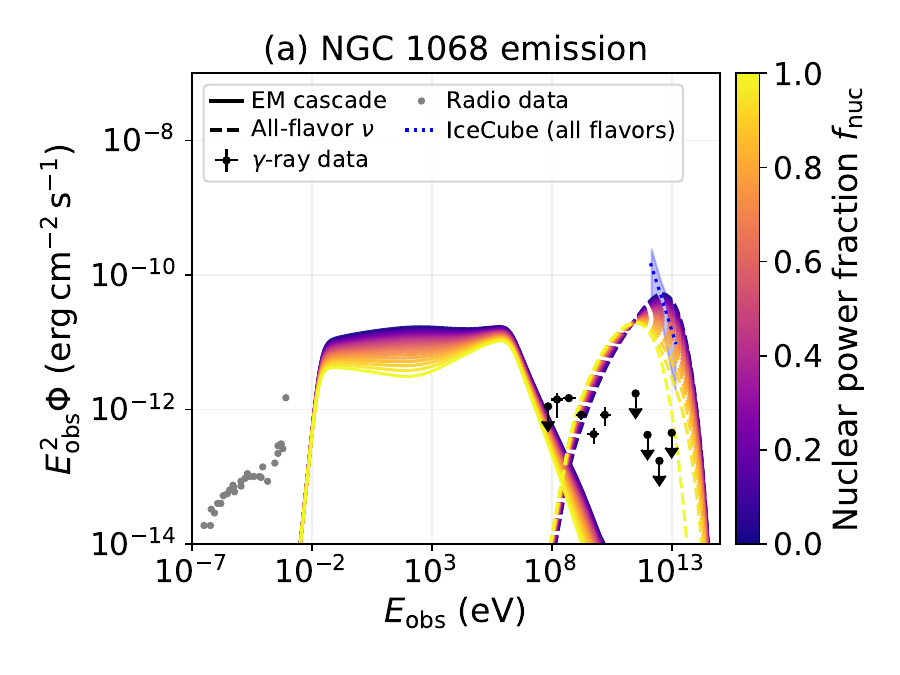}
\end{minipage}\hfill
\begin{minipage}[t]{0.5\textwidth}
\vspace{0pt}
\centering
\includegraphics[width=\linewidth]{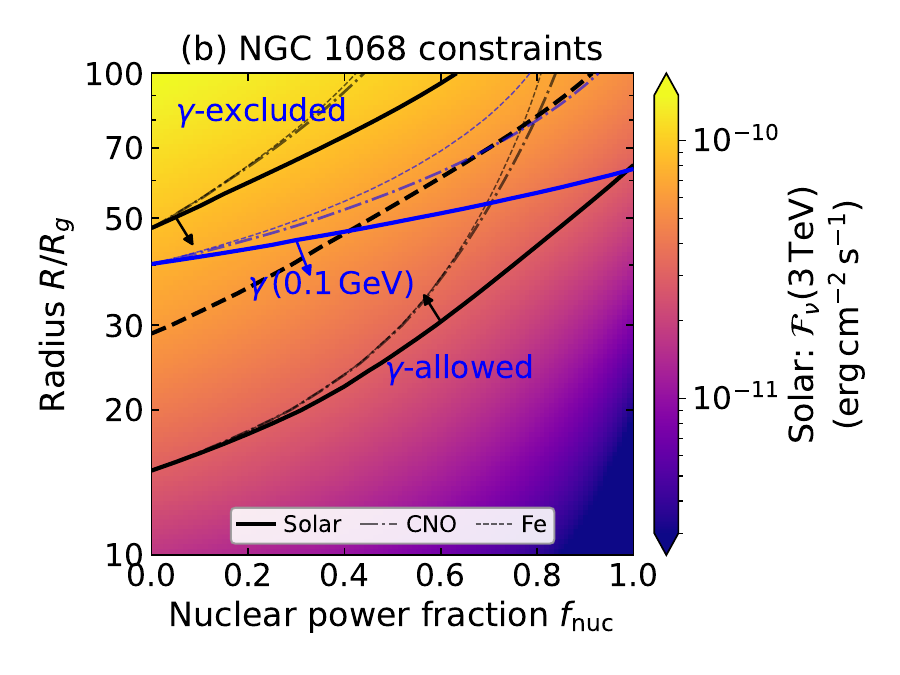}
\end{minipage}
\caption{{Multimessenger spectra and parameter selections for NGC~1068.
(a) Observer-frame spectra for the Solar template at $R=30R_g$, colored by $f_{\rm nuc}$; solid and dashed curves show secondary photons and all-flavor neutrinos.
The blue curve and bands show neutrino observations; black and gray points show gamma-ray and radio data.
(b) Colors show $\mathcal F_\nu(3\,\mathrm{TeV})$ for the Solar template.
Solid, dash-dotted, and thin dashed contours denote Solar, CNO, and Fe, respectively; black and blue indicate neutrino bounds and the gamma-ray ceiling.
For Solar, the thick black dashed curve marks $\mathcal F_\nu=6.5\times10^{-11}\,\mathrm{erg\,cm^{-2}\,s^{-1}}$; black arrows point into the neutrino interval, and blue annotations indicate the gamma-ray selection.}}
\label{fig:multimessenger}
\end{figure*}

\section{Multimessenger results and constraints}
\label{sec:multimessenger}
Figure \ref{fig:multimessenger}(a) shows the neutrino and EM-cascade spectra for the Solar template at $R=30R_g$.
We compare with the {all-flavor} IceCube spectrum and archival EM data compiled in Ref.~\cite{IceCube2022}, including observations by the Fermi Large Area Telescope (Fermi-LAT)~\cite{Ballet2020FermiDR2} and the MAGIC telescopes~\cite{Acciari2019MAGIC}.
{We find that increasing} $f_{\rm nuc}$ weakens the HE neutrino tail and shifts the spectral peak to lower energies, while the lower-energy emission becomes relatively more prominent.
The photon flux near $0.1$~GeV also decreases.
At small $f_{\rm nuc}$, the $3$~TeV neutrino flux approximately follows $(1-f_{\rm nuc})$ times the pure-proton result.
Recent IceCube--DeepCore upper limits provide a complementary probe of the composition-dependent sub-TeV neutrino emission~\cite{IceCubeDeepCore2026}, but are not included in the flux selections below.

Figure~\ref{fig:multimessenger}(b) compares the three templates over $10\leq R/R_g\leq100$ and $0\leq f_{\rm nuc}\leq1$, keeping the remaining source parameters fixed.
We define the observer-frame spectral energy fluxes $\mathcal F_\nu(E_\nu)=E_\nu^2\Phi_{\nu,\rm all}(E_\nu)$ and $\mathcal F_\gamma(E_\gamma)=E_\gamma^2\Phi_{\gamma,\rm sec}(E_\gamma)$, where $\Phi_{\nu,\rm all}$ and $\Phi_{\gamma,\rm sec}$ are the differential number fluxes of all-flavor neutrinos and secondary photons, respectively.
For NGC~1068, these diagnostics are evaluated at neutrino energy $E_\nu=3$~TeV and photon energy $E_\gamma=0.1$~GeV.
The neutrino diagnostic lies near the energy where the relative width of the IceCube spectral band is smallest~\cite{IceCube2022}.
The adopted selections are $3.0\times10^{-11}\leq\mathcal F_\nu\leq9.4\times10^{-11}$ and $\mathcal F_\gamma\leq10^{-12}$, with fluxes in $\mathrm{erg\,cm^{-2}\,s^{-1}}$.
The photon ceiling limits the coronal contribution to the measured gamma-ray emission~\cite{Ajello2023HadronicComponents}.
{The scan also reflects the spectral changes with $f_{\rm nuc}$ shown in Fig.~\ref{fig:multimessenger}(a).}

{The combined changes in available power, acceleration, escape, and target densities make the neutrino lower threshold favor larger radii, while the gamma-ray ceiling limits the radius from above.
The Solar template permits nuclear fractions approaching unity over a narrow radius interval, whereas CNO and Fe yield similar upper bounds around $f_{\rm nuc}\sim0.8$.}

{These template-dependent constraints can be understood by separating the neutrino contributions from proton and nuclear injection.}
At fixed source parameters and total CR power, the neutrino flux approximately satisfies
\begin{equation}
\frac{\mathcal F_\nu}{\mathcal F_{\nu,p}}\simeq(1-f_{\rm nuc})+f_{\rm nuc}r_\nu,
\label{eq:neutrino-yield}
\end{equation}
where $\mathcal F_{\nu,p}$ is the pure-proton flux and $r_\nu\equiv\mathcal F_\nu(f_{\rm nuc}=1)/\mathcal F_{\nu,p}\lesssim1$ is the corresponding ratio for purely nuclear injection, including all secondary contributions.
The supplied particle spectra and subsequent nuclear transport determine $r_\nu$, which depends on neutrino energy, source parameters, and the injected mixture.
In the relevant radius range, the helium-dominated Solar template has a larger $r_\nu$ at $3$~TeV than CNO or Fe, more effectively compensating for reduced primary-proton power.
The neutrino lower threshold therefore permits larger $f_{\rm nuc}$, subject also to the template-dependent gamma-ray ceiling.

At fixed black-hole mass and magnetization, our prescription gives $t_{\rm acc}^{-1}\propto(\eta R)^{-1}$, $L_{\rm CR}\propto\epsilon_{\rm CR}R/\eta$, and hence $L_p\propto(1-f_{\rm nuc})\epsilon_{\rm CR}R/\eta$.
{These scalings imply that the inferred bounds on $f_{\rm nuc}$ can shift: changes in available power rescale the emission, whereas acceleration and transport also modify the nuclear yield $r_\nu$.
The available power further depends on assumptions about coronal magnetization and confinement~\cite{Inoue2024CoronalBudget}.
Nevertheless, parameter combinations that reproduce the observed neutrino spectrum and satisfy the gamma-ray limits should supply a compatible range of parent-nucleon energies, constraining the relevant CR cutoff scales within the adopted transport model.
Since fragmentation approximately preserves energy per nucleon, different parameter combinations yielding similar nuclear cutoff energies per nucleon can retain similar limitations on the high-energy neutrino contribution from nuclei.}

\begin{figure}[t]
\includegraphics[width=\columnwidth]{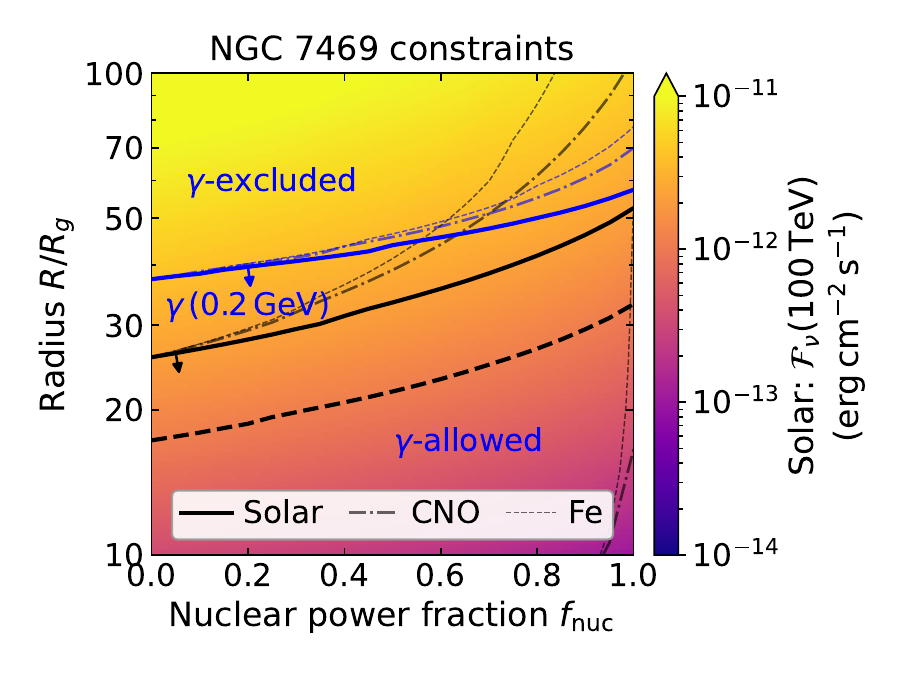}
\caption{{Parameter selections for NGC~7469, following the conventions of Fig.~\ref{fig:multimessenger}(b).
For the Solar template, colors show $\mathcal F_\nu(100\,\mathrm{TeV})$, and the thick black dashed curve marks $1.1\times10^{-12}\,\mathrm{erg\,cm^{-2}\,s^{-1}}$.
The adopted neutrino interval and gamma-ray ceiling at $0.2$~GeV are given in the text.}}
\label{fig:ngc7469-scan}
\end{figure}

\section{NGC~7469 at higher neutrino energies}
\label{sec:ngc7469}
We apply the same coronal framework to NGC~7469, whose harder candidate neutrino spectrum probes higher parent-nucleon energies~{\cite{IceCube2026AGN,Sommani2025NGC7469}}.
We adopt $M_{\rm BH}=9\times10^6M_\odot$, $d_L=71$~Mpc, and $z=0.016$~{\cite{Prince2025NGC7469Echo}}, and fix $(\eta,\epsilon_{\rm CR})=(0.005,0.02)$ while varying radius and composition.
The target radiation uses the broadband template adopted in Ref.~\cite{Testagrossa2026CoronalTurbulence}, normalized to the band luminosity $L_{2\text{--}10\,\mathrm{keV}}=1.45\times10^{43}\,\mathrm{erg\,s^{-1}}$~{\cite{Prince2025NGC7469Echo}}, using the coronal volume for the full template: $n_\gamma(\epsilon)=(R/c)\mathcal L_{\rm template}(\epsilon)/(V\epsilon^2)$.
Here $\mathcal L_{\rm template}$ is the template luminosity per logarithmic photon-energy interval.

Under these conditions, the supplied nuclear spectra extend to higher energies than in the NGC~1068 benchmark, and nuclear transport populates a broader range of intermediate isotopes (Figs.~\ref{fig:sm-ngc7469-density} and \ref{fig:sm-ngc7469-cascade}).
The higher parent energy per nucleon allows fragmentation to supply more energetic secondary nucleons.
The {HE neutrino peak for the Solar template} decreases and its highest-energy tail weakens as $f_{\rm nuc}$ increases, although substantial emission remains even at $f_{\rm nuc}=1$ (Fig.~\ref{fig:sm-ngc7469-sed}).

Figure~\ref{fig:ngc7469-scan} compares the three templates using adopted flux selections informed by the IceCube spectrum and Fermi-LAT upper limits~\cite{IceCube2026AGN,Eichmann2026SeyfertStarburst}.
We require $10^{-14}\leq\mathcal F_\nu(100\,\mathrm{TeV})\leq2\times10^{-12}$ and $\mathcal F_\gamma(0.2\,\mathrm{GeV})\leq3.0\times10^{-14}$, in $\mathrm{erg\,cm^{-2}\,s^{-1}}$.
The broad adopted neutrino interval reflects the large observational uncertainty and weakens the constraint on $f_{\rm nuc}$.
Large $f_{\rm nuc}$ remains possible for all three templates over part of the radius range, extending the allowed CNO and Fe fractions beyond those in NGC~1068.
Both neutrino and gamma-ray boundaries separate at large $f_{\rm nuc}$, with the neutrino upper-flux boundary shifting to larger radii for the heavier templates.
{For CNO and Fe, gamma-ray limits can set the maximum allowed radius at large $f_{\rm nuc}$.}
The numerical bounds remain source-specific because both coronal conditions and observational selections differ.

\section{Conclusions and outlook}
\label{sec:conclusions}
{Our calculations show that the neutrino yield at a given nuclear power fraction depends on the injected mixture and the energy per nucleon supplied by acceleration, with alternative momentum-diffusion scalings yielding steeper high-energy Fe tails (Fig.~\ref{fig:sm-dpp}).}

For NGC~1068, the joint flux selections yield template-dependent bounds on the nuclear power fraction: the helium-dominated Solar template permits larger $f_{\rm nuc}$ than CNO or Fe.
The adopted NGC~7469 model yields higher nuclear energies and a broader cascade, while its source-specific flux selections allow larger CNO and Fe fractions than in NGC~1068.
For the Solar template in both sources, increasing $f_{\rm nuc}$ changes the neutrino spectral shape and weakens its highest-energy tail, while appreciable emission survives at $f_{\rm nuc}=1$.

{Candidate excesses from NGC~4151 and CGCG~420-015~\cite{IceCube2025Seyfert}, with best-fit neutrino indices near $2.7$~\cite{IceCube2026AGN}, motivate applying this framework with source-specific conditions and constraints.
The collective excess from southern Seyferts motivates population studies~\cite{IceCube2026SouthernSeyfert}.} For the TeV--PeV CRs relevant here, magnetic deflections hinder a direct association with individual Seyferts~\cite{HAWCIceCube2019Anisotropy}.
Extragalactic magnetic diffusion may further suppress the arriving flux through a rigidity-dependent magnetic horizon~\cite{MollerachRoulet2019Diffusion}.
Neutrino and EM observations thus provide more direct probes of coronal acceleration and constraints on the power assigned to nuclei.
{Future population studies of Seyfert AGNs could combine the escaping nuclear spectra with propagation models to assess their contribution to the observed CR spectrum and composition.}

\begin{acknowledgments}
The author thanks Walter Winter for helpful comments and suggestions. ChatGPT-6 Astra (OpenAI) was used to improve the language and readability, with all changes reviewed and validated by the author. This work is supported in part by IISN project No.~4.4501.18.
\end{acknowledgments}

\onecolumngrid
\appendix

\section{Acceleration, transport and interaction details}
\label{app:transport}

\subsection{Particle acceleration}
\label{app:particle-acceleration}

{We describe turbulent particle acceleration in the coronae of active galactic nuclei (AGNs) using the Fokker--Planck (FP) equation}
\begin{align}
\frac{\partial f_i}{\partial t}
={}\frac{1}{p^2}\frac{\partial}{\partial p}
\left[
 p^2D_{pp,i}\frac{\partial f_i}{\partial p}
 +p^2b_{p,i}f_i
\right]+q_i^{\rm seed}
-\left(t_{{\rm esc},i}^{-1}+\Gamma_i^{\rm cat}\right)f_i.
\label{eq:sm-fp}
\end{align}
For species $i$, $f_i(p,t)$ is the isotropic phase-space density in particle momentum $p$ and time $t$, with number density $n_i$ defined by $dn_i=4\pi p^2f_i\,dp$.
{The coefficient $b_{p,i}=-\dot p_i>0$ describes continuous momentum losses, and $t_{{\rm esc},i}$ is the spatial escape time.
The rate $\Gamma_i^{\rm cat}$ describes the loss of primary species $i$ through destructive interactions or proton-to-neutron conversion and enters the sink term $-\Gamma_i^{\rm cat}f_i$.
The injection term $q_i^{\rm seed}$ supplies seed particles in phase space.}
Each primary is evolved independently in the prescribed coronal background~\cite{Yuan2026CoupledAcceleration}.

We use the momentum-diffusion coefficient $D_{pp,i}=p^2/t_{\rm acc}$, with the acceleration time defined in the main text.
{Following Ref.~\cite{Yuan2026CoupledAcceleration}, we adopt a seed spectrum with characteristic momentum $p_{{\rm pk},i}$,}
\begin{equation}
q_i^{\rm seed}(p)\propto
\frac{p}{p_{{\rm pk},i}}
\exp\!\left(-\frac{p}{p_{{\rm pk},i}}\right),
\qquad
\frac{p_{{\rm pk},i}}{m_ic}\simeq5.0.
\label{eq:sm-seeds}
\end{equation}
{At momenta well above the seed scale, the shape of the accelerated spectra is insensitive to the detailed seed spectrum.}
Spatial escape is described by
\begin{align}
r_{L,i}=\frac{pc}{|Z_i|eB},
\qquad
\lambda_i=l_c\left(\frac{r_{L,i}}{l_c}\right)^{1/3},
\qquad
t_{{\rm esc},i}=
\max\left(\frac{R^2}{c\lambda_i},\frac Rc\right).
\label{eq:sm-escape}
\end{align}
Here $r_{L,i}$ is the Larmor radius, $\lambda_i$ the scattering mean free path with a rigidity scaling motivated by Ref.~\cite{Lemoine2023SharpBends}, $Z_i$ the charge number, and $e$ the elementary charge; $l_c$ is the coherence length defined in the main text.

The continuous energy-loss coefficient $b_{E,i}$ includes synchrotron radiation, Bethe--Heitler (BH) pair production~\cite{Chodorowski1992Photopair}, proton photomeson interactions and interactions with ambient protons.
{For the latter, we adopt an effective fractional energy-loss rate $b_{E,i}^{ip}/E=\kappa_{ip}\Gamma_{ip}$, where $\Gamma_{ip}$ is the collision rate with ambient protons.
We set $\kappa_{ip}=0.5$ for all species as a phenomenological extension of the usual $pp$ approximation~\cite{Murase2020HiddenCores}.
For nuclei, this prescription does not explicitly resolve the energy partition among pions, outgoing nucleons, and residual fragments.}
Proton photomeson interactions are treated with the factorized response-function approach of Ref.~\cite{Hummer2010Photohadronic,Klinger2024AM3}, including resonance, direct, and multipion production.
{Continuous cooling accounts for energy losses in channels retaining the proton, while proton-to-neutron conversion contributes a separate sink term to the proton equation.}
{For nuclei, photomeson and photodisintegration interactions contribute to the loss rate $\Gamma_i^{\rm cat}$ of the primary species and constrain the accelerated primary spectra.
Daughter-nucleus and secondary-nucleon production is omitted from the FP stage and treated in the subsequent nuclear cascade.}
Continuous energy and momentum losses are related by $b_{p,i}=b_{E,i}/v_i$, where $v_i=dE/dp$ is the particle speed.

\begin{figure*}[t]
\includegraphics[width=0.49\textwidth]{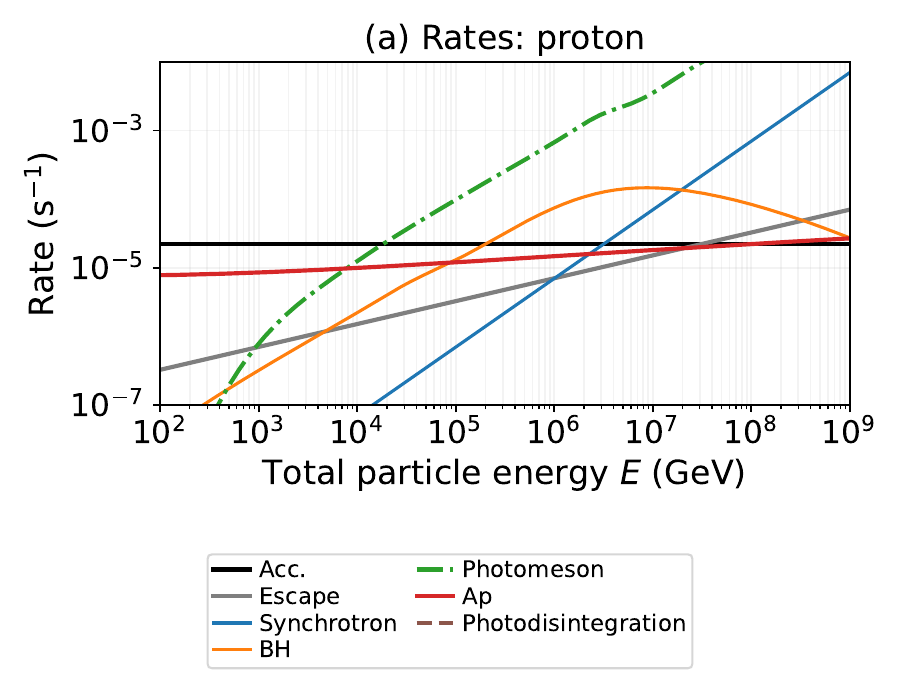}
\hfill
\includegraphics[width=0.49\textwidth]{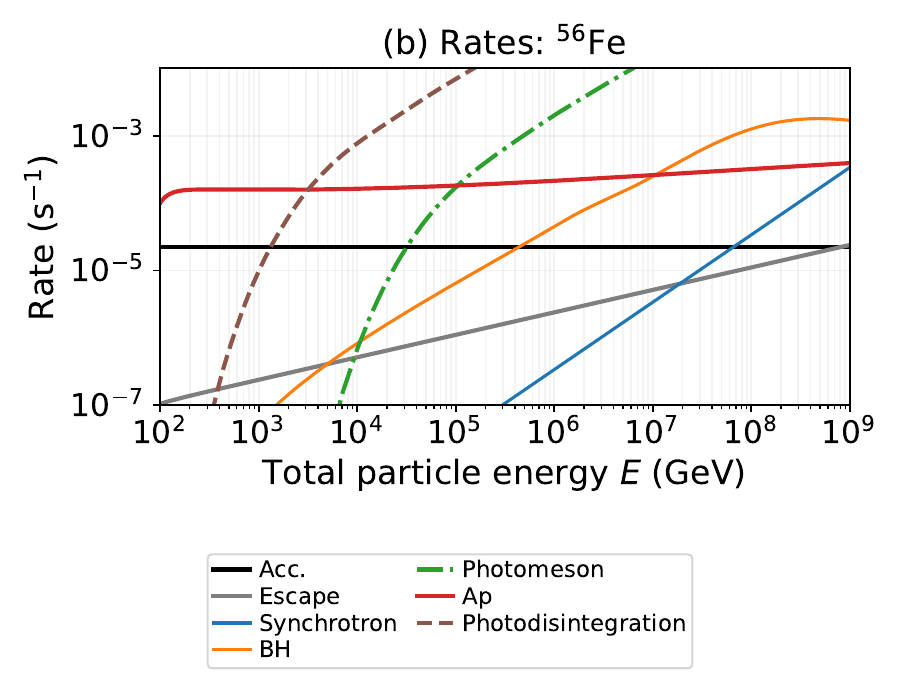}
\caption{{Acceleration, escape, interaction, and loss rates for protons (a) and ${}^{56}\mathrm{Fe}$ (b) in the NGC~1068 benchmark with $R=30R_g$, $f_{\rm nuc}=0.6$, and the Solar composition template.
The acceleration curve is $D_{pp}/p^2=t_{\rm acc}^{-1}$, with $t_{\rm acc}=4.5\times10^4$~s.
Synchrotron, Bethe--Heitler (BH), and $Ap$ curves show fractional continuous energy-loss rates, with $Ap$ denoting collisions with ambient protons (including $pp$).
Photodisintegration and iron photomeson curves show interaction rates.
The proton photomeson curve combines fractional cooling and proton-to-neutron conversion rates.}}
\label{fig:sm-primary-rates}
\end{figure*}

Figure~\ref{fig:sm-primary-rates} compares the acceleration, escape, interaction, and loss rates of protons and iron in the NGC~1068 benchmark used for the main-text cosmic-ray (CR) density figure.
{The plotted acceleration rate is $D_{pp}/p^2=t_{\rm acc}^{-1}$.}
Iron already undergoes substantial gas-collision cooling before photodisintegration rises rapidly at higher energies.
{The cutoff shape is determined by the full FP solution, including spatial escape and sink terms, and cannot be assigned to a single rate crossing.}

{The stationary FP solution supplies the spectral shape of the effective injection, $Q_i^{\rm eff}(E)=C_iN_i^{\rm FP}(E)$. Following Eq.~(\ref{eq:effective-injection}), we fix $C_i$ through}
\begin{equation}
{V\int dE\,T_i(E)Q_i^{\rm eff}(E)=L_i.}
\label{eq:sm-injection-power}
\end{equation}
{Here $T_i(E)=E-m_ic^2$ is the particle kinetic energy, and the differential number density per unit energy is}
\begin{equation}
N_i^{\rm FP}(E)=\frac{4\pi p^2}{v_i}f_i(p).
\label{eq:sm-energy-density}
\end{equation}
{We retain the particle speed in this conversion, including near the rest-energy threshold.}

We solve the stationary equation on a logarithmic momentum grid with zero flux at the lower boundary and an absorbing upper boundary.
Solver checks include particle balance and agreement between stationary solutions and late-time evolution for prescribed transport coefficients.

\begin{figure}[t]
\includegraphics[width=0.6\columnwidth]{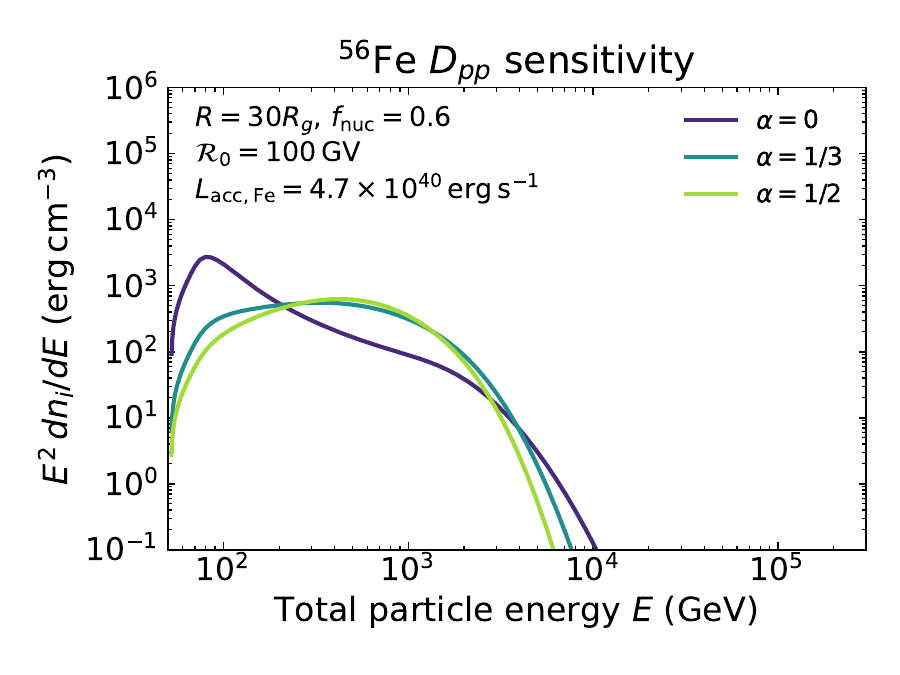}
\caption{{Stationary FP spectra of surviving Fe for the momentum-diffusion prescriptions in Eq.~(\ref{eq:sm-dpp-test}), using the NGC~1068 benchmark with $R=30R_g$, $f_{\rm nuc}=0.6$, and the Solar template.
The mean momentum gain rate is matched at $100$~GV, and the effective Fe injection power is held fixed.
Daughter nuclei are not evolved.}}
\label{fig:sm-dpp}
\end{figure}

{We test the momentum dependence of acceleration for ${}^{56}\mathrm{Fe}$ in the NGC~1068 {benchmark with the Solar template}, keeping the cooling, interaction rates, spatial escape, and seed shape fixed.}
{For rigidity $\mathcal R=pc/(|Z|e)$, we parameterize the momentum dependence as}
\begin{equation}
{\frac{p^2}{D_{pp}}
=t_{\rm acc}\frac{4-\alpha}{4}
\left(\frac{\mathcal R}{\mathcal R_0}\right)^\alpha,}
\label{eq:sm-dpp-test}
\end{equation}
with reference rigidity $\mathcal R_0=100$~GV and rigidity-scaling index $\alpha=0,1/3,1/2$, corresponding to $D_{pp}\propto p^2,p^{5/3},p^{3/2}$.
The alternative scalings $D_{pp}\propto p^{5/3}$ and $p^{3/2}$ are motivated by gyroresonant acceleration in Kolmogorov and Kraichnan turbulence, respectively~\cite{BeckerLeDermer2006}.
{The factor $(4-\alpha)/4$ ensures that the mean fractional momentum gain rate, $(4-\alpha)D_{pp}/p^2$, matches the baseline value $4/t_{\rm acc}$ at $\mathcal R_0$.}
{The effective injection associated with each spectrum is normalized to the same Fe power, $L_{\rm Fe}=4.69\times10^{40}\,\mathrm{erg\,s^{-1}}$, using Eq.~(\ref{eq:effective-injection}).}
Figure~\ref{fig:sm-dpp} shows that the alternatives reduce the low-energy accumulation, enhance the intermediate-energy population, and steepen the highest-energy tail.
{This single-species FP comparison tests the sensitivity of the supplied Fe spectrum to the momentum dependence of $D_{pp}$. The main-text results retain the baseline prescription $\alpha=0$, corresponding to $D_{pp}\propto p^2$.}

\subsection{Nuclear interactions and cascades}
\label{app:nuclear-interactions}

{Using the effective injection derived from the FP spectra, we follow the subsequent nuclear cascade in the corona, tracking primary nuclei, daughter nuclei, and secondary nucleons.}
The time-dependent differential number density per unit energy $N_i(E,t)$ of each nuclear species satisfies
\begin{align}
\frac{\partial N_i}{\partial t}
={}&Q_i^{\rm eff}(E)
+\frac{\partial}{\partial E}\!\left[b_{E,i}(E)N_i(E,t)\right]
\nonumber\\
&-\left[t_{{\rm esc},i}^{-1}(E)+\Gamma_i^{\rm rem}(E)\right]N_i(E,t)
\nonumber\\
&+\sum_j\int dE'\,
\mathcal K_{j\rightarrow i}(E',E)N_j(E',t).
\label{eq:sm-cascade}
\end{align}
For the stationary injection and coronal conditions considered here, we solve Eq.~(\ref{eq:sm-cascade}) with $\partial N_i/\partial t=0$ and denote the resulting distribution by $N_i(E)$.
{Here $b_{E,i}$ contains continuous energy losses, while $\Gamma_i^{\rm rem}$ is the loss rate of species $i$ due to photonuclear interactions, gas collisions, and decay, entering the sink term $-\Gamma_i^{\rm rem}N_i$.}
{The production kernel $\mathcal K_{j\rightarrow i}(E',E)$ gives the differential production rate of species $i$ at energy $E$ per parent particle of species $j$ at energy $E'$, including product multiplicities.}
{It accounts for fragmentation, photomeson interactions, and nuclear decay.}
{The kernels are constructed from the interaction and decay data described below, using a treatment similar to that of Ref.~\cite{Biehl2018NuclearCascade}.}
{Charged fragments escape on the timescale given by Eq.~(\ref{eq:sm-escape}), while neutrons escape on the timescale $R/c$.}
Fragmentation is treated in the approximation that the daughters retain the parent Lorentz factor, so $E_d/A_d\simeq E_j/A_j$ for a relativistic parent $j$ and daughter $d$.
No momentum-diffusion term acts on these secondaries in Eq.~(\ref{eq:sm-cascade}); fragmentation can populate lighter species but cannot itself raise their energy per nucleon above that supplied by the parent.

{For an isotropic target photon field, the photonuclear interaction rate is}
\begin{align}
\Gamma_{i\gamma}(E)={}
\frac{c}{2\gamma_i^2}
\int_{\bar\epsilon_{\rm th}}^\infty
 d\bar\epsilon\,
 \bar\epsilon\,\sigma_i(\bar\epsilon)\int_{\bar\epsilon/(2\gamma_i)}^\infty
 d\epsilon\,\frac{n_\gamma(\epsilon)}{\epsilon^2},
\label{eq:sm-photonuclear-rate}
\end{align}
Here $\gamma_i$ is the Lorentz factor of species $i$, and $\sigma_i$ is the cross section for the interaction considered.
Barred energies refer to the nuclear rest frame, with $\bar\epsilon_{\rm th}$ denoting the threshold photon energy.
{For photodisintegration, this rate gives the corresponding contribution to $\Gamma_i^{\rm rem}$.}
Energy-loss rates use the corresponding inelasticity-weighted responses.

The nuclear data are assembled into a common isotope and reaction bundle extending to $A=90$.
Nuclear masses and decay branches are obtained through the \textsc{CRPropa} data preparation framework~\cite{AlvesBatista2016CRPropa3} using inputs from the National Institute of Standards and Technology (NIST)~\cite{NISTAtomicWeights} and the NuDat3 database~\cite{NNDCNuDat3}.
Photonuclear parent nuclei are selected with rest-frame mean lifetimes exceeding $0.1$~s; daughter and decay products complete the isotope inventory.
Below $200$~MeV in the nuclear rest frame, we use the evaluated light-nucleus cross sections distributed with \textsc{CRPropa} for supported $A<12$ targets~\cite{Kampert2013CRPropa2} and calculations with the \textsc{talys} nuclear-reaction code~\cite{Koning2023TALYS} for $A\geq12$.
{We obtain the production cross section for each daughter species by summing over all reaction channels, weighted by the number of particles of that species produced. The products include residual nuclei and emitted neutrons, protons, deuterons, tritons, ${}^{3}\mathrm{He}$, and ${}^{4}\mathrm{He}$.}
From $0.2$ to $1$~GeV, we use the empirical nuclear-fragmentation component of the \textsc{AstroPhoMes} library~\cite{Morejon2019AstroPhoMes} for total photonuclear and inclusive secondary-nucleon and fragment cross sections.
The fragmentation cross sections are truncated immediately above $1$~GeV; the zero point at $2$~GeV is retained to represent the integrated-response tail.
{These inclusive production cross sections determine the daughter-production rates through Eq.~(\ref{eq:sm-photonuclear-rate}), while the retained-Lorentz-factor approximation specifies the daughter energies.}
Only species reachable from the injected primaries are included in the transport network.

Photomeson production is calculated separately using nucleon responses from the \textsc{sophia} event generator~\cite{Mucke2000SOPHIA} in the superposition approximation:
\begin{align}
\Gamma_{A\gamma}^{\rm mes}(E)={}
Z\Gamma_{p\gamma}^{\text{\textsc{sophia}}}(E/A)
+(A-Z)\Gamma_{n\gamma}^{\text{\textsc{sophia}}}(E/A).
\label{eq:sm-photomeson}
\end{align}
The corresponding responses provide the outgoing nucleon and pion spectra.
The fragmentation and \textsc{sophia}-based photomeson terms enter the transport equation additively, without event-level matching of reaction channels.
A bound-nucleon interaction also supplies the residual nucleus.
When a residual is absent from the isotope table, it is replaced by an admissible known nucleus and excess free nucleons, preserving mass number and charge.
This prescription supplies an approximate network closure rather than microscopic residual branching information.
For gas collisions, the $pp$ cross section and the SIBYLL-based inclusive pion parametrization follow Ref.~\cite{KelnerAharonianBugayov2006}.
{For nuclear collisions with ambient protons, we adopt $\sigma_{Ap}(E)=A^{3/4}\sigma_{pp}(E/A)$~\cite{JoshiWinterGupta2014}.}
{The pion injection rates are summed over the transported species, using the same inclusive distribution for $\pi^+$, $\pi^-$, and $\pi^0$.}
Collisions with ambient protons remove the projectile and produce pions; outgoing baryon and residual-fragment regeneration from these collisions are omitted.

\begin{figure}[t]
\includegraphics[width=0.6\columnwidth]{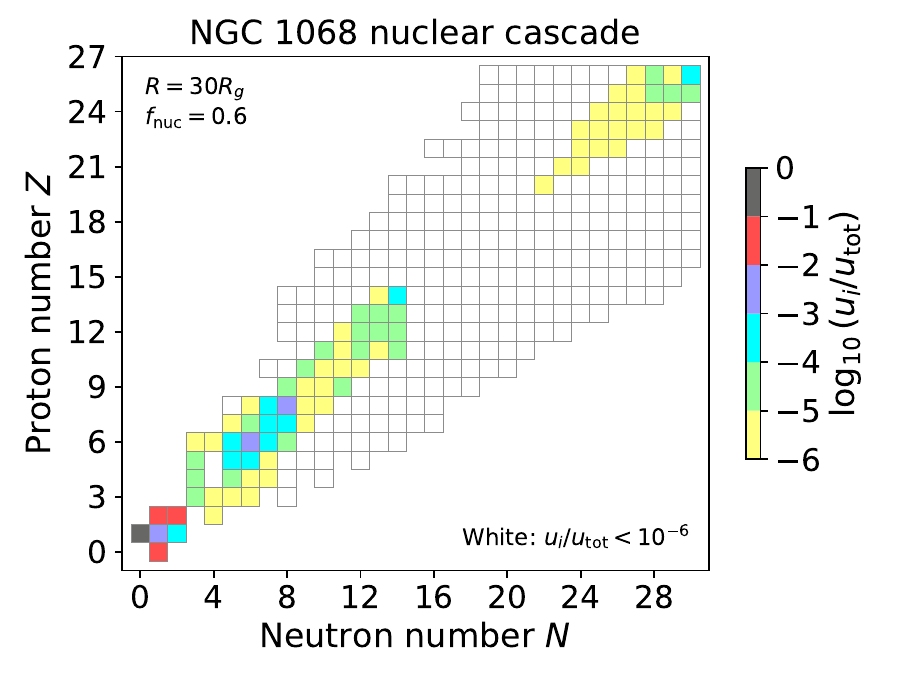}
\caption{{Transported nuclear composition for the NGC~1068 benchmark ($R=30R_g$, $f_{\rm nuc}=0.6$, Solar template).
Cells represent isotopes with neutron and proton numbers $(N,Z)$.
Colors show $\log_{10}(u_i/u_{\rm tot})$, where $u_i=\int_{100\,\mathrm{GeV}}^\infty E N_i(E)\,dE$ is the stored energy density and $E$ the total particle energy.
The sum $u_{\rm tot}=\sum_j u_j$ includes all transported nuclei and nucleons, including species below the $10^{-6}$ display threshold (white).}}
\label{fig:sm-cascade}
\end{figure}

{\textsc{NuCascade}, the nuclear-transport code developed and used here, solves the stationary energy equation for each species and iterates the secondary sources until the network meets its convergence criterion, followed by a source-consistency check.}
{Figure~\ref{fig:sm-cascade} shows the nuclear composition after transport for the NGC~1068 benchmark with $R=30R_g$, $f_{\rm nuc}=0.6$, and the Solar template. It illustrates how nuclear fragmentation redistributes the injected energy among surviving nuclei, lighter fragments, and secondary nucleons. The color scale indicates each species' contribution to the total stored CR energy above a total particle energy of $100$~GeV.}

{The pion injection rates per unit volume and energy obtained from \textsc{NuCascade},}
\begin{equation}
Q_{\pi^a}(E)=Q_{\pi^a}^{\,\gamma}(E)+Q_{\pi^a}^{\,{\rm gas}}(E),
\quad a\in\{+,-,0\},
\label{eq:sm-pion-sources}
\end{equation}
are transferred to the \textsc{am}$^3$ radiation code~\cite{Klinger2024AM3} before pion cooling or decay.
The superscripts $\gamma$ and ${\rm gas}$ denote production in photomeson interactions and collisions with ambient protons, respectively.
{Electron--positron pairs produced by BH interactions of protons and nuclei are also injected into \textsc{am}$^3$.}

{We approximate nuclear BH pair production using the proton differential kernel at the same Lorentz factor, with coherent $Z^2$ scaling. Taking the nuclear mass to be $A m_p$, we map the steady nuclear distributions $N_{A,Z}(E_A)=dn_{A,Z}/dE_A$ onto an effective proton distribution,}
\begin{equation}
{N_{p,\mathrm{eff}}(E_p)=\sum_{A,Z} A Z^2 N_{A,Z}(A E_p).}
\label{eq:sm-nuclear-bh-pairs}
\end{equation}
{Here $E_p=E_A/A$, and the factor $A$ accounts for $dE_A=A\,dE_p$. Convolving this distribution with the \textsc{am}$^3$ proton BH kernel~\cite{Klinger2024AM3,KelnerAharonian2008Photohadronic} and the coronal photon field yields equal electron and positron injection spectra, which are evolved in the electromagnetic cascade. We check the resulting pair-injection power against the nuclear BH energy-loss power without renormalizing the pair spectra. This approximation neglects finite nuclear size and recoil corrections.}

The radiation calculation includes pion and muon decay together with electromagnetic (EM) cascades, treating synchrotron and inverse-Compton processes of charged secondaries, photon--photon pair production, synchrotron self-absorption, and escape.
The nuclear-cascade proton density additionally supplies proton synchrotron emission; pion production is evaluated only in the nuclear-transport stage.
External photons are continuously supplied at the injection rate per unit volume and energy $Q_{\gamma,\rm ext}=n_\gamma/(R/c)$, and the evolving radiation is not returned to the acceleration or nuclear calculations.

We also neglect
electromagnetic injection from nuclear beta-decay electrons and
de-excitation photons, while retaining beta decay in the nuclear
transport. For ultrarelativistic parent nuclei, the mean secondary
energy fractions are
$\zeta_j\equiv\langle E_j\rangle/E_A
\simeq\langle\bar E_j\rangle/(A m_p c^2)$, with $j=e,\gamma$.
For iron-group nuclei and MeV-scale rest-frame energies, these
fractions are of order $10^{-5}$ per emitted particle, motivating
their treatment as subdominant radiative channels. 

Verification includes analytic escape and cooling limits and comparisons of coupled nuclear evolution with independent integration.
The radiation calculation uses a time step of $10^{-3}R/c$ and an evolution interval of $15R/c$ to ensure convergence.

\begin{figure}[tp]
\includegraphics[width=0.6\columnwidth]{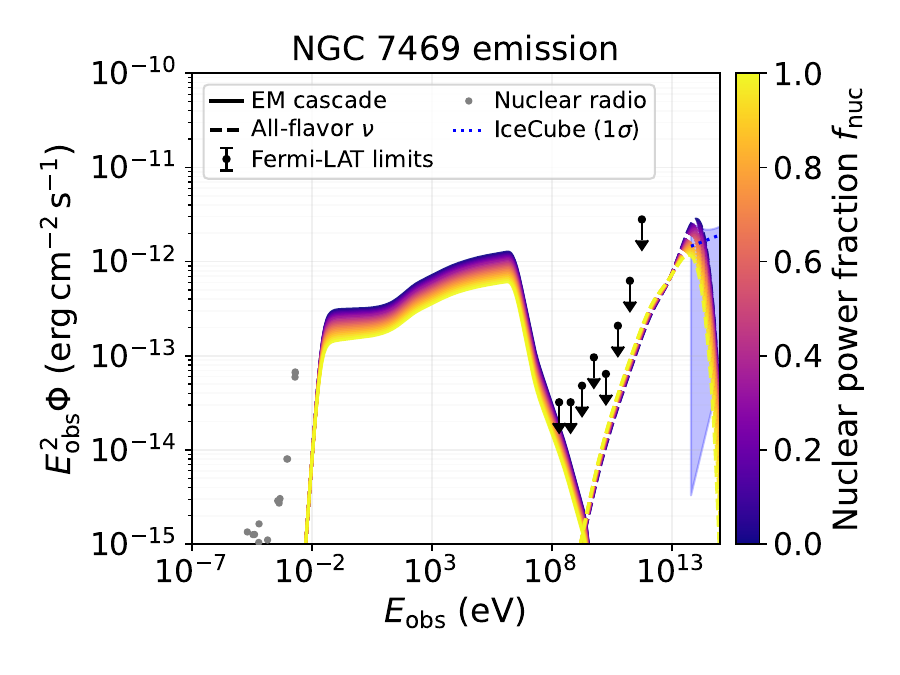}
\caption{{Predicted NGC~7469 spectra for the Solar template at $R=30R_g$ and varying $f_{\rm nuc}$.
Solid and dashed curves show secondary photons and all-flavor neutrinos, respectively.
The blue curve and band show the adopted neutrino reference and IceCube uncertainty, respectively; black arrows are Fermi-LAT upper limits, and gray points are archival nuclear radio measurements.}}
\label{fig:sm-ngc7469-sed}
\end{figure}

\section{Nuclear cascades and multimessenger emission in NGC 7469}
\label{app:ngc7469}

{Figure~\ref{fig:sm-ngc7469-sed} shows the multimessenger spectra of NGC~7469 for the Solar template at $R=30R_g$ as $f_{\rm nuc}$ varies.}
Increasing $f_{\rm nuc}$ lowers the high-energy (HE) neutrino peak and weakens its highest-energy tail.
The accompanying EM cascade also decreases, while substantial neutrino emission from the Solar template remains at $f_{\rm nuc}=1$.
{The blue curve is an adopted neutrino reference spectrum using the rounded power-law parameters compiled in Ref.~\cite{Testagrossa2026CoronalTurbulence}. The IceCube band is the published all-flavor $1\sigma$ flux envelope over the reported sensitive energy interval~\cite{IceCube2026AGN}.}
The Fermi Large Area Telescope (Fermi-LAT) points are the $95\%$ spectral upper limits of Ref.~\cite{Eichmann2026SeyfertStarburst}.
The adopted flux selections in the main text are conditional diagnostics, rather than a fit to independent spectral bins.
Archival nuclear radio measurements~\cite{Prieto2010NuclearSED,Eichmann2026SeyfertStarburst} are shown for reference and are not treated as isolated coronal detections.

\begin{figure}[t]
\includegraphics[width=0.6\columnwidth]{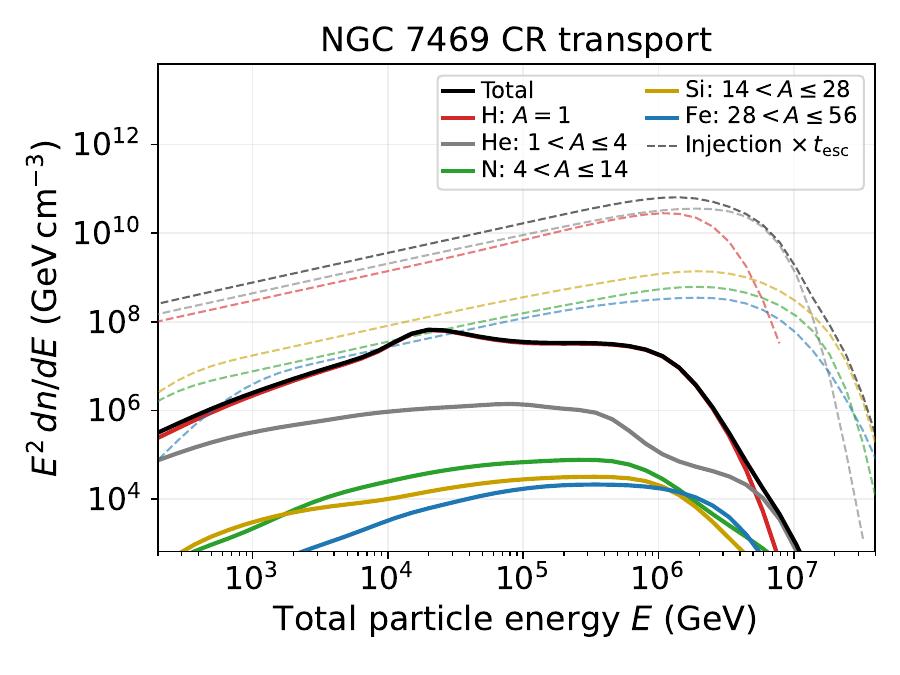}
\caption{{CR densities for the NGC~7469 Solar template at $R=30R_g$ and $f_{\rm nuc}=0.6$.
Solid and dashed curves show stationary $E^2N_i(E)$ after transport and the injection--escape reference $E^2Q_i^{\rm eff}t_{{\rm esc},i}$, respectively, summed over mass groups.
H includes protons and neutrons; black curves show totals.}}
\label{fig:sm-ngc7469-density}
\end{figure}

Figure~\ref{fig:sm-ngc7469-density} compares the transported spectra with the injection--escape reference, which is not the FP density itself.
Compared with the NGC~1068 benchmark in the main text, the supplied HE tails extend further, while the surviving heavy groups are strongly depleted during transport.
The H group includes primary protons and secondary protons and neutrons, so it does not isolate the contribution from nuclear fragmentation.

\begin{figure}[t]
\includegraphics[width=0.6\columnwidth]{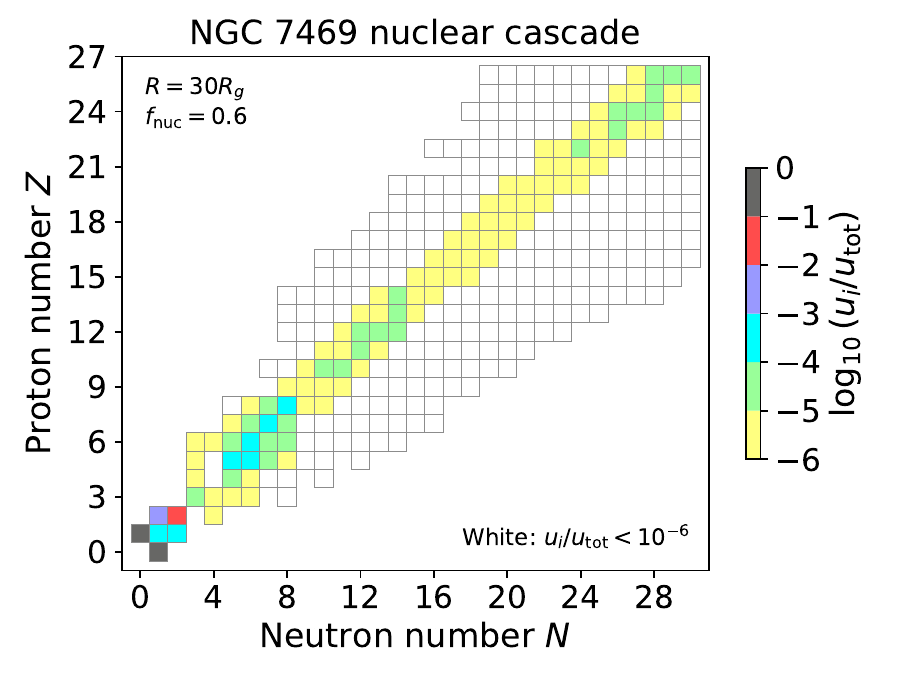}
\caption{Nuclear composition after transport in the NGC~7469 {benchmark with $R=30R_g$, $f_{\rm nuc}=0.6$, and the Solar template}.
Conventions are as in Fig.~\ref{fig:sm-cascade}.}
\label{fig:sm-ngc7469-cascade}
\end{figure}

At the common benchmark values $R=30R_g$ and $f_{\rm nuc}=0.6$, the NGC~7469 isotope distribution in Fig.~\ref{fig:sm-ngc7469-cascade} contains a more continuous sequence of intermediate nuclei above the display threshold than the NGC~1068 benchmark in Fig.~\ref{fig:sm-cascade}.
The two source benchmarks have different radiation fields and acceleration parameters, so this comparison does not isolate a single changed parameter.
The common $100$~GeV total-energy cut corresponds to different energy-per-nucleon thresholds for different isotopes.
Moreover, the {Solar template} directly injects several nuclei, and the stationary distributions do not retain ancestry labels.
The charts therefore illustrate the populated network but do not measure the fraction of Fe undergoing complete breakup or the nuclear share of the neutrino emission.

\clearpage
\twocolumngrid

\bibliography{ref}

\end{document}